\documentclass[twocolumn,superscriptaddress,amsmath,ammsymb]{revtex4-1}
\usepackage{epsfig}
\usepackage{amsmath}
\usepackage{amssymb}
\usepackage{graphicx}
\usepackage{dcolumn}
\usepackage{bm}
\usepackage[usenames]{color}
\usepackage[breaklinks]{hyperref}
\usepackage{soul}
\usepackage{ulem}
\hypersetup{colorlinks=true,allcolors=blue}

\begin{document}
	
\title{Magnetotransport and activation energy of the surface states in Cd$_3$As$_2$ thin films}

\author{Zhigang Cai}
\affiliation{School of Science, Jiangnan University, Wuxi 214122, China} 
\author{Fuxiang Li}
\email{fuxiangli@hnu.edu.cn}
\affiliation{School of Physics and Electronics, Hunan University, Changsha 410082, China}
\author{Yi-Xiang Wang}
\email{wangyixiang@jiangnan.edu.cn}
\affiliation{School of Science, Jiangnan University, Wuxi 214122, China}
\affiliation{School of Physics and Electronics, Hunan University, Changsha 410082, China}

\date{\today}

\begin{abstract} 
Recent experiments performed the magnetotransport measurements in (001)-oriented  Cd$_3$As$_2$ thin films and attributed the magnetotransport properties to the surface states.  
In this paper, by using an effective model to describe the surface states, we analyze the Landau bands and then calculate the magnetoconductivities and magnetoresistivities.  
From these results, the features of two-dimensional quantum Hall effect of the surface states can be captured. 
More importantly, we reveal that the activation energy is determined by the Hall plateau width, which can explain the experimental observations that the activation energies at odd plateaus are larger than those at even plateaus.  We also analyze the roles played by the structural inversion symmetry breaking and impurity scatterings in the magnetotransport, and suggest that their combined effects would lead to the absence of some Hall plateaus.    
\end{abstract} 

\maketitle

\section{Introduction} 

The field of condensed matter physics has witnessed the ongoing interests and developments of  topological materials~\cite{M.Z.Hasan, X.L.Qi}.  Among them, three-dimensional (3D) Dirac semimetals (DSMs) belong to the phases of matter with gapless excitations that act as 3D counterparts of two-dimensional (2D) graphene~\cite{N.P.Armitage, B.Q.Lv}.  An important feature is that 3D DSMs can host Fermi arcs that connect the bulk node projections on the surface Brillouin zone (BZ)~\cite{X.G.Wan, A.C.Potter, P.J.W.Moll}.  Besides that, due to the intrinsic band inversions, 3D DSMs also support the surface states that resemble those of 3D topological insulators (TIs)~\cite{M.Kargarian, B.J.Yang}.  When the thickness of a 3D DSM is reduced to below a critical value, the exotic physical properties dominated by the surface states will be manifested due to the quantum confinement, for which Cd$_3$As$_2$ serves as an example.  

The bulk Cd$_3$As$_2$ represents a prototype 3D DSM~\cite{Z.Wang, S.Borisenko, T.Liang, Z.K.Liu, S.Jeon} and its thin film configuration was predicted to be a nearly ideal wide-gap 2D TI with a high surface state mobility~\cite{Z.Wang}.  Thanks to the improvements of fabrication techniques and sample quality, Cd$_3$As$_2$ thin films of several tens of nanometers have been successfully realized in recent experiments~\cite{B.Guo, D.A.Kealhofer,A.C.Lygo}.  The magnetotransport measurements performed in (001)-oriented Cd$_3$As$_2$ thin films at low temperatures reported the quantized Hall resistivity $\rho_{xy}$ and the concomitant almost vanishing longitudinal resistivity $\rho_{xx}$~\cite{B.Guo, D.A.Kealhofer, A.C.Lygo}, which provide clear evidences for 2D quantum Hall effect (QHE).  
These results strongly suggested that the surface states were responsible for the transport signatures, whereas the bulk state contributions were excluded.  Besides that, the topic of QHE has been extensively investigated in Cd$_3$As$_2$, including the quantum Hall state variation with the film thickness~\cite{M.Uchida}, the transport mechanism based on the surface states in (112)-oriented thin film~\cite{T.Schumann} and on the Weyl orbits in wedge-shaped nanostructures~\cite{C.Zhang}. 

For the finite-temperature magnetotransport in Cd$_3$As$_2$ thin films~\cite{D.A.Kealhofer}, the Hall plateau $\nu=1$ was found to be abruptly preempted by an insulating state at a larger gate bias, which is accompanied by the collapse of the QHE. 
The absence of some Hall plateaus was also observed and was explained by the structural inversion symmetry (SIS) breaking.   
Moreover, the characteristic activation energies that are connected to the dips of the longitudinal resistivity were found to be larger at odd plateaus than those at even plateaus.  In their paper~\cite{D.A.Kealhofer}, the authors attributed the activation energy difference to the intersurface Coulomb interactions and pointed out that such interactions could drive the spontaneous coherent state between the top and bottom surface states, which was called as a quantum Hall superfluid or an exciton condensate~\cite{J.P.Eisenstein, D.Tilahun}.  However, the quantum Hall superfluid behavior was suggested to occur only in the TI thin films less than 30 nm, as delimited by the dielectric constant of the materials~\cite{D.Tilahun}; while the Cd$_3$As$_2$ thin film samples measured in the experiments were 45 nm and 50 nm thick, which far exceed 30 nm.  We also note that odd plateaus and even plateaus are just labels that are introduced to distinguish the different Hall plateaus; they are not supposed to be favored by interactions, but should be equally affected.  Thus, it would be interesting to explore the physics underlying the activation energy difference. 

Motivated by these progresses, in this paper, we will study the magnetotransport properties of the surface states in (001)-oriented Cd$_3$As$_2$ thin films theoretically, to explore the features of 2D QHE as well as the activation energy behavior.  We will adopt an effective model to describe the surface states and analyze the Landau bands under the magnetic field.  Then we use the Kubo-Bastin formula to calculate the magnetoconductivities and magnetoresistivities, from which the activation energy can be extracted. 
We further analyze the roles played by the SIS breaking and impurity scatterings in the magnetotransport. 
The SIS breaking was induced by the potential offset between the two spatially separated surfaces~\cite{Y.Zhang, W.Y.Shan}, as they were connected to the different substrates.  Experimentally, the SIS breaking was believed to exist in thicker Cd$_3$As$_2$ films with thickness $d>45$ nm~\cite{D.A.Kealhofer}, but not in thinner films with $d<22$ nm~\cite{A.C.Lygo}.  Moreover, the impurity scatterings play an indispensable role in understanding the transport behavior of topological systems, e.g., 3D QHE~\cite{Y.X.Wang2023} and anomalous Hall effect~\cite{H.W.Wang} in 3D topological material ZrTe$_5$. 

Our main results are given as follows: (i) we capture the features of 2D QHE of the surface states, including the peaks in the longitudinal conductivity $\sigma_{xx}$ and quantized plateaus in the Hall conductivity $\sigma_{xy}$, which agree well with the experiments~\cite{B.Guo, D.A.Kealhofer, A.C.Lygo};  
(ii) we reveal that the activation energy is determined by the Hall plateau width, which could explain the activation energy difference between odd and even plateaus within the noninteracting single-particle picture;  
(iii) we attribute the Hall plateaus absence to the combined effects of the SIS breaking and impurity scatterings. 
Our work could better help understand the magnetotransport behaviors in Cd$_3$As$_2$ thin film experiments.

\section{Model and LLs}

\begin{figure}
	\includegraphics[width=9.2cm]{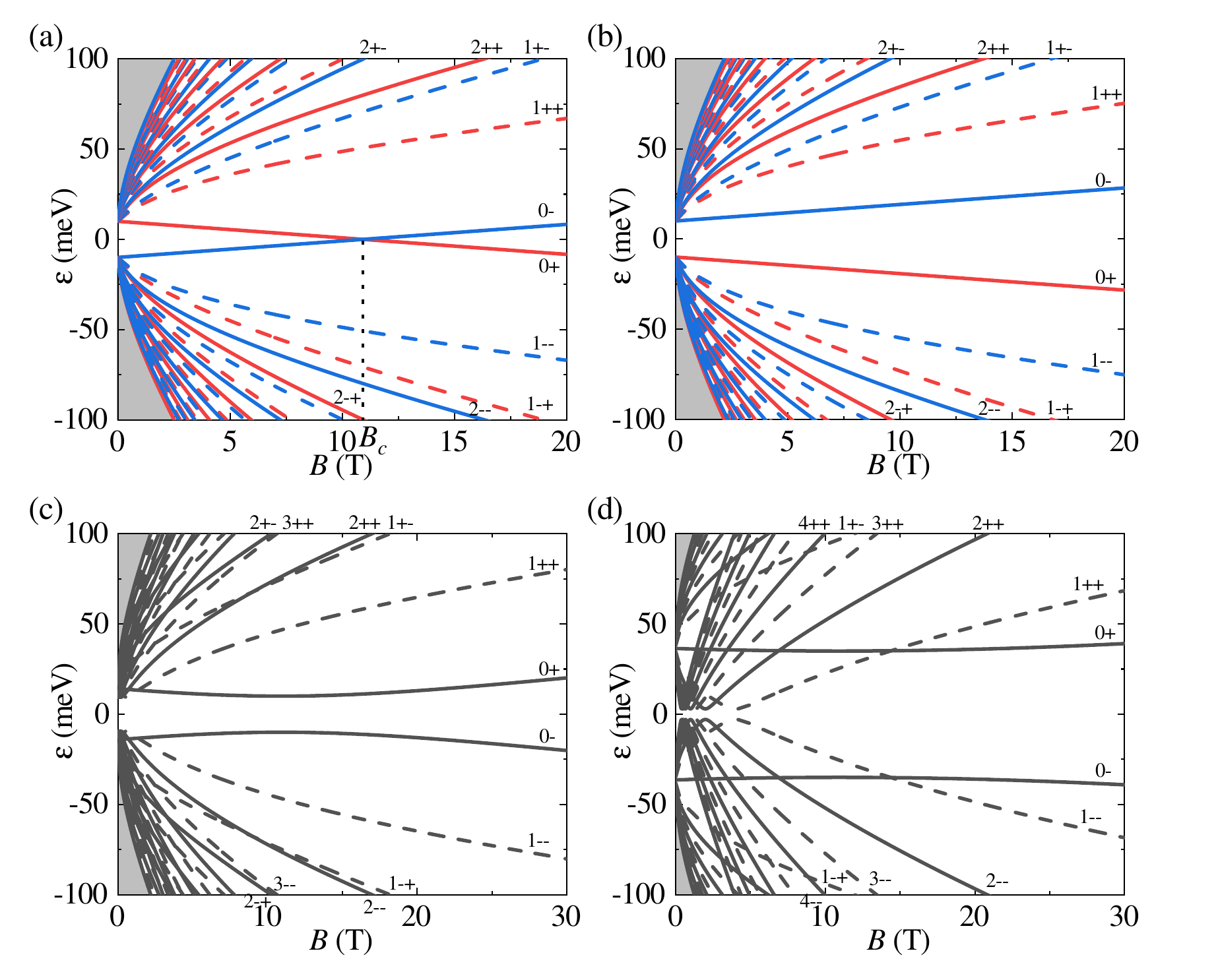}
	\caption{(Color online) The LL spectra of Cd$_3$As$_2$ thin films, with the index  labeled as $(ns\lambda)$.  The even (odd) parity is represented by the solid (dashed) line and the top (bottom) surface state is denoted by the red (blue) line.  We choose $v=5\times10^5$ m/s, $\xi=600$ meV nm$^2$, and the other model parameters as: (a) $V_0=0$, $\Delta=20$ meV; (b) $V_0=0$, $\Delta=-20$ meV; (c) $V_0=10$ meV,  $\Delta=20$ meV; and (d) $V_0=35$ meV, $\Delta=20$ meV.  Note that in (c) and (d), as the top and bottom surface states are hybridized, the black lines are used. }
	\label{Fig1}
\end{figure} 

In the surface BZ of a DSM, the bulk states and surface states will coincide unless a gap is opened  in the bulk~\cite{X.Xiao}, which can be achieved by the quantum confinement and symmetry breaking in a thin film configuration~\cite{T.Schumann, C.Zhang}.  In (001)-oriented Cd$_3$As$_2$ thin films, the observed magnetotransport behaviors are well explained by the effective model describing the low-energy surface states around the $\Gamma$ point~\cite{D.A.Kealhofer, B.Guo, A.C.Lygo}, which demonstrate the existence of the surface states.  Such an effective model is written as ($\hbar=1$)~\cite{H.Z.Lu},
\begin{align}
&\hat H(\boldsymbol k)=\begin{pmatrix}
\hat h_+(\boldsymbol k)& V_0 I_2
\\
V_0 I_2& \hat h_-(\boldsymbol k)  
\end{pmatrix}, 
\end{align}
with
\begin{align} 
\hat h_{\tau_z}(\boldsymbol k)
=-v(k_x\sigma_y+k_y\sigma_x)+\tau_z\big(\frac{\Delta}{2}-\xi k^2\big)\sigma_z. 
\end{align}
Here $\sigma$ and $\tau$ are the Pauli matrices and act on the upspin/downspin and top/bottom surface, respectively.  
$v$ is the Fermi velocity, $\Delta$ is the Dirac mass, $\xi$ is the coefficient of the quadratic term on the diagonal. 
$V_0$ denotes the SIS breaking parameter.  When $V_0=0$, the system owns the SIS as $\hat{\cal I}^{-1}\hat H(\boldsymbol k)\hat{\cal I}=\hat H(-\boldsymbol k)$, with the operator $\hat{\cal I}=\tau_z\otimes I$; the introduction of the $V_0$ term will break the SIS.  

Under a uniform magnetic field $\boldsymbol B=B\boldsymbol e_z$, the one-dimensional Landau bands will be formed.  To solve the Landau bands, we use the Peierls substitution to change the kinetic momentum as the canonical momentum $\boldsymbol\pi=\boldsymbol k+e\boldsymbol A$, with the vector potential chosen as $\boldsymbol A=-By{\boldsymbol e}_x$ in the Landau gauge.  Then the canonical momenta are replaced by the ladder operators, 
$\pi_-=\pi_x-i\pi_y\rightarrow\frac{\sqrt2}{l_B}\hat a$ and $\pi_+=\pi_x+i\pi_y\rightarrow \frac{\sqrt2}{l_B}\hat a^\dagger$, in which $[\hat a,\hat a^\dagger]=1$ and the magnetic length $l_B=\frac{1}{\sqrt{eB}}$.  With the trial wavefunction $\psi_n=(c_{n1}|n\rangle,c_{n2}|n-1\rangle,c_{n3}|n\rangle,c_{n4}|n-1\rangle)^T$, the problem of calculating the LL energies, $H_{nB}|\psi_n\rangle=\varepsilon_n|\psi_n\rangle$,  is reduced to finding the eigenvalues of 
\begin{align}
&H_{nB}
\nonumber
\\
&=\begin{pmatrix}
\frac{\Delta-\omega}{2}-n\omega& 
i\frac{\sqrt{2n}v}{l_B}& V_0& 0
\\
-i\frac{\sqrt{2n}v}{l_B}& -\frac{\Delta+\omega}{2}+n\omega& 0& V_0
\\
V_0& 0& -\frac{\Delta-\omega}{2}+n\omega& i\frac{\sqrt{2n}v}{l_B}
\\
0& V_0& -i\frac{\sqrt{2n}v}{l_B}& \frac{\Delta+\omega}{2}-n\omega 
\end{pmatrix}. 
\end{align}
The energies for the $n=0$ and $n\geq1$ LLs are solved as 
\begin{align}
&\varepsilon_{0s}=s\Big[\frac{(\Delta-\omega)^2}{4}+V_0^2\Big]^\frac{1}{2},
\label{varepsilon0}
\\
&\varepsilon_{ns\lambda}
=s\Big[\Big(\varepsilon_n-s\lambda\sqrt{\frac{\omega^2}{4} 
+\frac{2nv^2V_0^2}{l_B^2\varepsilon_n^2}}\Big)^2
+V_0^2 \frac{(\Delta-2n\omega)^2}{4\varepsilon_n^2}\Big]^\frac{1}{2}, 
\label{varepsilonn}
\end{align}
respectively, where $\varepsilon_n=\sqrt{\frac{(\Delta-2n\omega)^2}{4}+\frac{2nv^2}{l_B^2}}$ and $\omega=\frac{2\xi}{l_B^2}$.  The index $s=\pm1$ denotes the conduction/valence band, and $\lambda=\pm1$ characterizes the two branches that represent the top/bottom surface state when $V_0=0$.  Since the system owns the parity symmetry with the operator $\hat{\cal P}=(-1)^{\hat a^\dagger \hat a} I\otimes \sigma_z$, each LL carries a definite even or odd parity~\cite{T.Devakul, Y.X.Wang2022}.  
The LL spectra are plotted in Fig.~\ref{Fig1}, with the even (odd) parity represented by the solid (dashed) lines, and top (bottom) surface states by the red (blue) lines. 

First we consider the $V_0=0$ case that there is no SIS breaking.  In Fig.~\ref{Fig1}(a) when the Dirac mass $\Delta>0$, we see that the magnetic field will drive the gap closing and reopen, with the zeroth LLs crossing at the critical field $B_c$, which represents a hallmark of 2D TI under a magnetic field~\cite{M.Z.Hasan, X.L.Qi}.  In the experiment~\cite{A.C.Lygo}, the zeroth LL crossings are clearly captured and the critical field $B_c$ decreases with reducing thickness $d$ in the range (18-22) nm.  From Eq.~(\ref{varepsilon0}), we have $B_c=\frac{\Delta}{2e\xi}$, thus the Dirac mass $\Delta$ also decreases with reducing $d$.  In Fig.~\ref{Fig1}(b) when $\Delta<0$, the gap widens persistently with the magnetic field and the zeroth LLs will not cross, meaning that the system behaves as a topologically trivial insulator.  This is consistent with the observations in $d=14$ nm Cd$_3$As$_2$ thin film~\cite{A.C.Lygo}. 
Therefore through changing the film thickness, the Dirac mass $\Delta$ can be effectively  modulated and even reverse its sign, leading to the transition from the topologically trivial phase to the nontrivial phase.  Compared to 2D TIs that are realized in HgTe/CdTe quantum well~\cite{B.A.Bernevig2006b, M.Konig} and monolayer transition metal dichalcogenides WTe$_2$~\cite{S.Wu, S.Tang}, here changing the thickness of topological semimetal thin films can open a new direction to realize such a long-sought phase~\cite{C.L.Kane, C.C.Liu, M.Ezawa, X.Qian, M.A.Cazalilla}. 

\begin{figure}
	\includegraphics[width=9.2cm]{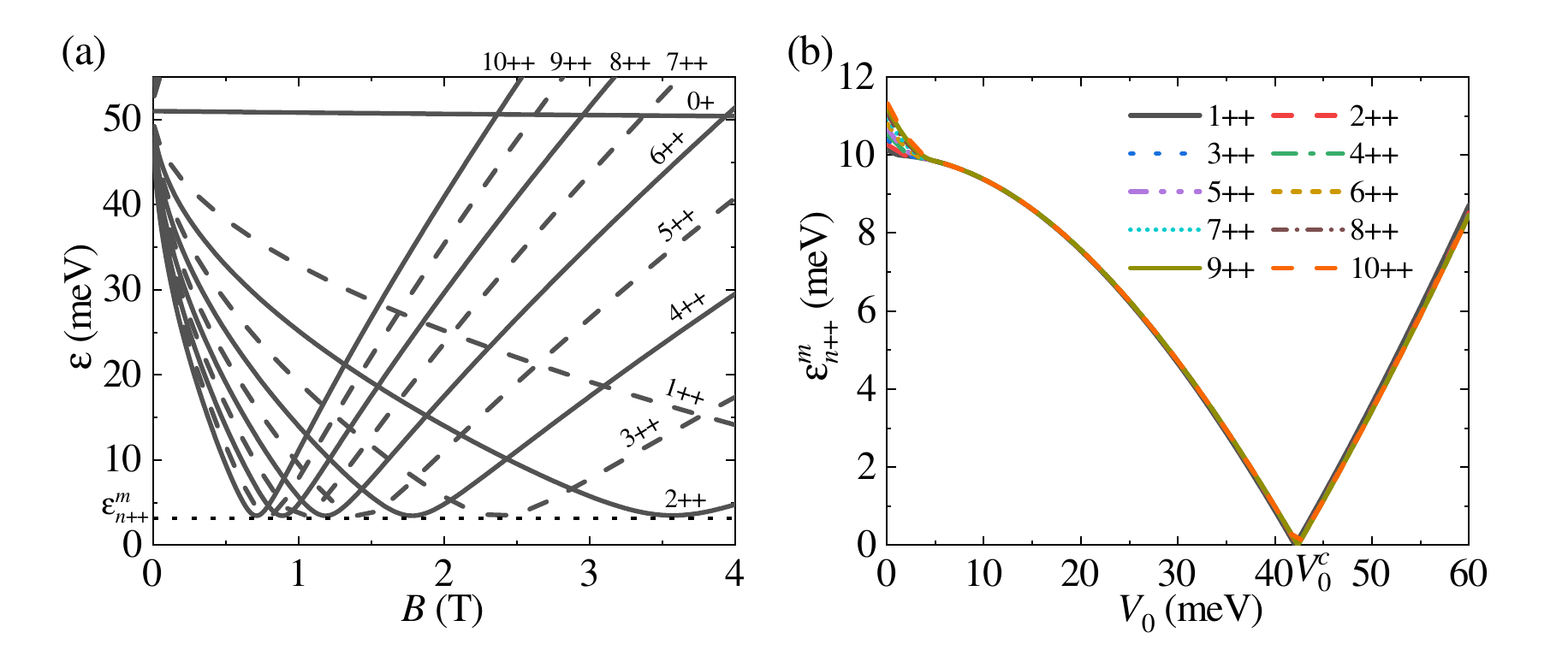}
	\caption{(Color online) (a) The enlarged LL spectra of Fig.~\ref{Fig1}(d) under a weak magnetic field $B<4$ T.  (b) The minimum energy $\varepsilon_{n++}^m$ of the LL branch $(n++)$ versus $V_0$.  At the critical $V_0^c$, the gap is closed for all LL branches.  
	The model parameters are chosen as $v=5\times10^5$ m/s, $\xi=600$ meV nm$^2$, $V_0=35$ meV, and $\Delta=20$ meV.} 
	\label{Fig2}
\end{figure}

For a finite $V_0$, the SIS breaking will hybridize the top and bottom surface states, but the parity property of the LLs is still preserved.  In Fig.~\ref{Fig1}(c) with $V_0=10$ meV and $\Delta=20$ meV, we see that (i) the zeroth LLs become anti-crossing and is accompanied by the nonmonotonic gap evolution; (ii) the $n\geq1$ LLs get crossed at higher energies, which is vital in explaining the Hall plateau absence, as discussed below.  Further increasing $V_0$ to the strong $V_0=35$ meV in Fig.~\ref{Fig1}(d), the LL crossings at higher energies are more evident.  We observe that the zeroth LLs remain almost unchanged to the magnetic field.  By contrast, the $(n++)$ LL branch decreases at low fields and is followed by an increasing, in which the magnetic field at the transition point grows with the index $n$.  This is more clearly seen in the enlarged LL spectra for $B<4$ T in Fig.~\ref{Fig2}(a).  

To illustrate the evolutions of the $(n++)$ LL branch with $V_0$, we locate its minimum energy $\varepsilon_{n++}^m$ and display the results in Fig.~\ref{Fig2}(b).  When $V_0>4$ meV, for different $n$, the minimum energies collapse on the same line and exhibit the same evolutions with $V_0$.  More importantly, the increasing $V_0$ can drive the gap closings at the critical $V_0^c=42.3$ meV and then reopen.  This means that a band inversion occurs, and correspondingly, there exists a topological phase transition. Actually, in Eq.~(\ref{varepsilonn}), by requiring $\varepsilon_{n++}=0$, the critical $V_0^c$ is solved as 
\begin{align}
V_0^c=\frac{v^2\Delta}{2\xi}-\frac{\Delta^2}{16 n^2}.  
\end{align}
The second term can be neglected as it is much smaller than the first one.  Thus, we obtain  $V_0^c=\frac{v^2\Delta}{2\xi}=42.36$ meV, which is independent of the index $n$ and agrees with the numerical results.  
The gap closings are understood from the dispersion that for a finite $V_0$, there is a competition between the SIS preserving term and SIS breaking term; when $V_0$ is large, the SIS breaking term becomes dominant and will drive the gap closings. 

As such a band inversion can occur in all $n\geq1$ LL branches, this is different from the above Dirac mass-driven band inversion that only occurs in the zeroth LLs.  Thus, the SIS breaking $V_0$ term provides another feasible route to modulate the topological phase transition in Cd$_3$As$_2$ thin films, which needs more studies in the future.

\section{magnetoconductivities and magnetoresistivities} 

Next we study the magnetotransport properties of Cd$_3$As$_2$ thin films.  By using the Kubo-Bastin formula~\cite{A.Bastin, L.Smrcka, G.D.Mahan}, the conductivity tensors $\sigma_{\alpha\beta}$ are expressed as functions of the chemical potential $\mu$, magnetic field $B$, and temperature $T$. 
\begin{align} 
&\sigma_{\alpha\beta}(\mu,B,T)=\frac{1}{2\pi A_0}\sum_{\boldsymbol k} 
\int_{-\infty}^\infty d\varepsilon f_T(\varepsilon)
\Big[\text{Tr}\big(\hat J_\alpha\frac{d\hat G^R}{d\varepsilon}
\nonumber\\
&\quad\quad\times\hat J_\beta (\hat G^A-\hat G^R)
-\hat J_\alpha(\hat G^A-\hat G^R) \hat J_\beta \frac{d\hat G^A}{d\varepsilon}
\big)\Big].  
\label{Kubo-Streda}
\end{align}
Here $A_0$ is the area of the 2D system, $f_T(\varepsilon)=(1+e^{\frac{\varepsilon-\mu}{k_BT}})^{-1}$ is the Fermi-Dirac distribution function with the Boltzmann constant $k_B$, $\hat J_\alpha=e\frac{\partial\hat H}{\partial k_\alpha}$ is the current density operator along the $\alpha$ direction, and $\hat G^{R/A}(\varepsilon,\eta)=(\varepsilon-\hat H\pm i\eta)^{-1}$ is the retarded/advanced Green's function, in which $\eta$ denotes the LL linewidth broadening that is introduced to represent the effect of the impurity scatterings phenomenologically~\cite{H.Watanabe2020, H.Watanabe2021}.  For simplicity, we  take $\eta$ as a constant for all LLs. 

\begin{figure}
	\includegraphics[width=9.2cm]{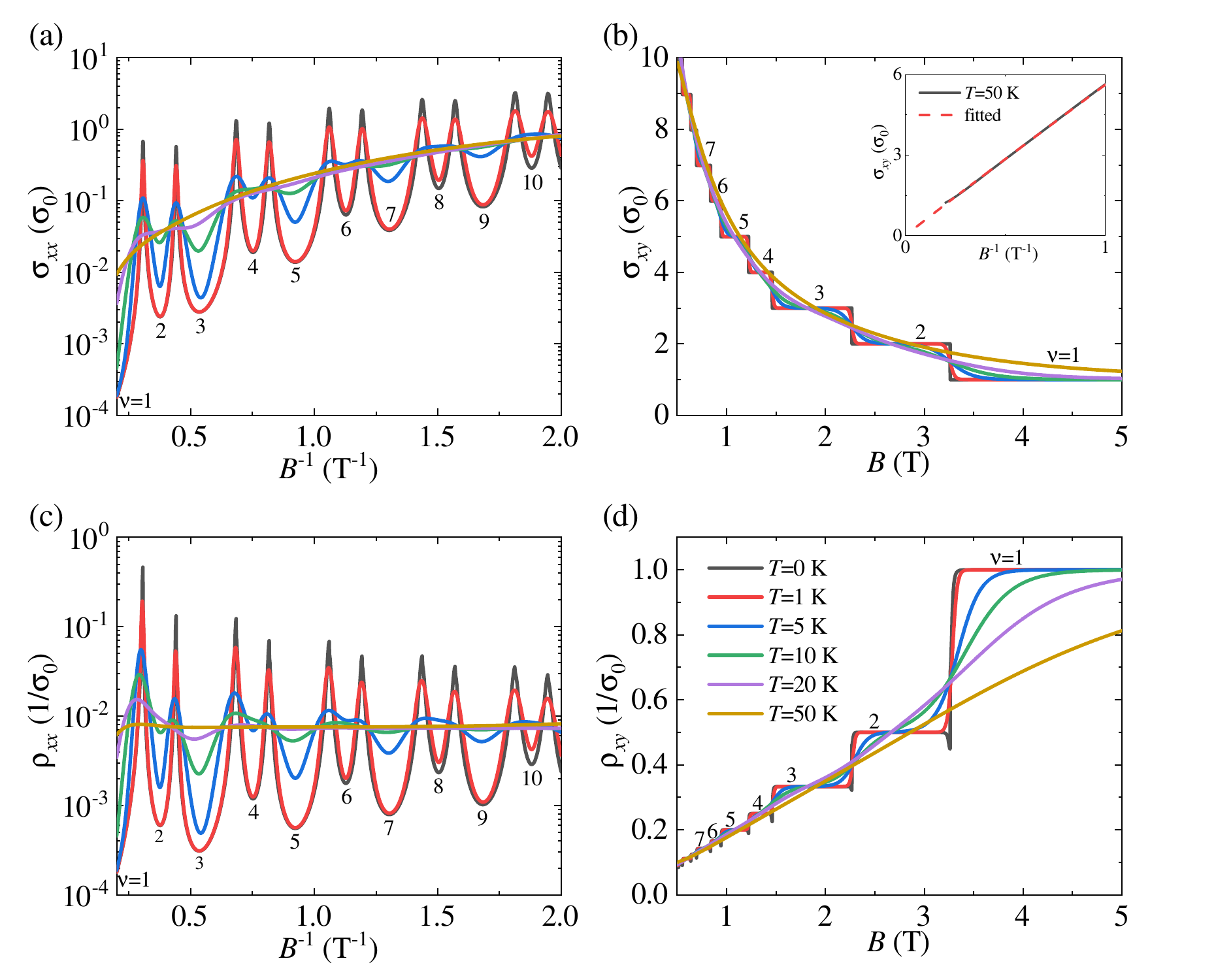}
	\caption{(Color online) Magnetoconductivities (a)$-$(b) and magnetoresistivities (c)$-$(d) of Cd$_3$As$_2$ thin films for different temperatures $T$.  The integer Hall plateaus $\nu$ are labeled and the legends are the same in all figures.  We choose the same model parameters as Fig.~\ref{Fig1}(a), and set the chemical potential $\mu=30$ meV and the linewidth $\eta=0.1$ meV.  In the inset of (b), $\sigma_{xy}$ at $T=50$ K is displayed versus the inverse magnetic field $B^{-1}$, which is fitted as $\sigma_{xy}=5.63 B^{-0.98}$.}
	\label{Fig3}
\end{figure}

With the help of the LL energies $\varepsilon_{ns\lambda}$ and wave functions $\psi_{ns\lambda}$, the zero-temperature longitudinal conductivity $\sigma_{xx}$ and Hall conductivity $\sigma_{xy}$ can be derived directly.  The expressions are given as 
\begin{align}
&\sigma_{xx}(\mu,B,T=0)=\sigma_0\frac{2\eta^2}{\pi l_B^2} 
\sum_{n\geq 0} \sum_{s,s'} \sum_{\lambda,\lambda'}  
\big|M_{ns\lambda;n+1,s'\lambda'}\big|^2
\nonumber\\
&\qquad\times
\frac{1}{[(\mu-\varepsilon_{ns\lambda})^2+\eta^2][(\mu-\varepsilon_{n+1,s'\lambda'})^2+\eta^2]},
\label{sigmaxx}
\end{align}
and 
\begin{align}
&\sigma_{xy}(\mu,B,T=0)=\sigma_0\frac{2}{l_B^2} 
\sum_{n\geq 0} \sum_{s,s'} \sum_{\lambda,\lambda'}  
\big|M_{ns\lambda;n+1,s'\lambda'}\big|^2
\nonumber\\
&\times\frac{(\varepsilon_{ns\lambda}-\varepsilon_{n+1,s'\lambda'})^2-\eta^2}
{[(\varepsilon_{ns\lambda}-\varepsilon_{n+1,s'\lambda'})^2+\eta^2]^2} \big[
\theta(\mu-\varepsilon_{ns\lambda})\theta(\varepsilon_{n+1,s'\lambda'}-\mu)
\nonumber\\
&-\theta(\mu-\varepsilon_{n+1,s'\lambda'})
\theta(\varepsilon_{ns\lambda}-\mu)\big], 
\label{sigmaxy} 
\end{align}
respectively.  Here $\sigma_0=\frac{e^2}{2\pi}=(25.8\text{ k}\Omega)^{-1}$ is the unit of the quantum conductivity, $\theta(x)$ is the step function, and the matrix element is 
\begin{align}
&M_{ns\lambda;n+1,s'\lambda'}
=iv(c_{ns\lambda}^{1*}c_{n+1,s'\lambda'}^2
+c_{ns\lambda}^{3*}c_{n+1,s'\lambda'}^4)
\nonumber
\\
&\quad\quad+\frac{\sqrt{2(n+1)}\xi}{l_B} 
(-c_{ns\lambda}^{1*}c_{n+1,s'\lambda'}^1+c_{ns\lambda}^{3*}c_{n+1,s'\lambda'}^3) 
\nonumber 
\\
&\quad\quad+\frac{\sqrt{2n}\xi}{l_B} (c_{ns\lambda}^{2*}c_{n+1,s'\lambda'}^2-c_{ns\lambda}^{4*}c_{n+1,s'\lambda'}^4).  
\end{align}
In Eqs.~(\ref{sigmaxx}) and (\ref{sigmaxy}), two aspects are worthy pointing out: (i) the nonvanishing matrix element of the current density determines the common selection rules $n\rightarrow n\pm1$; 
(ii) the conductivity components that are contributed by the LL transitions $(ns\lambda\rightarrow n's'\lambda')$ satisfy the following relations~\cite{Y.X.Wang2023, Z.Cai}
\begin{align}
&\sigma_{xx}(ns\lambda\rightarrow n+1,s'\lambda')
=\sigma_{xx}(n+1,s'\lambda'\rightarrow ns\lambda), 
\\
&\sigma_{xy}(ns\lambda\rightarrow n+1,s'\lambda')
=-\sigma_{xy}(n+1,s'\lambda'\rightarrow ns\lambda), 
\end{align}
which tells us that the contributions to $\sigma_{xx}$ from the LL transition $ns\lambda\rightarrow (n+1,s'\lambda')$ and from $(n+1,s'\lambda')\rightarrow ns\lambda$ are equal, while those to $\sigma_{xy}$ are opposite.  

At a finite temperature, $T$ will enter the conductivity through the Fermi-Dirac distribution function $f_T(\varepsilon)$.  By multiplying $\sigma_{\alpha\beta}$ with a temperature-dependent factor $-\frac{\partial f_T(\zeta-\mu)}{\partial\zeta}$, the finite-temperature conductivities can be expressed as a weighted integration of the zero-temperature conductivities around the chemical potential $\mu$~\cite{L.Smrcka}, 
\begin{align}
\sigma_{\alpha\beta}(\mu,B,T)=\int_{-\infty}^\infty d\zeta\sigma_{\alpha\beta}(\zeta,B,T=0)
\Big[-\frac{\partial f_T(\zeta-\mu)}{\partial\zeta}\Big].   
\label{finiteT}
\end{align} 
Note that at zero temperature, the temperature-dependent factor becomes a $\delta$-function, $-\frac{\partial f_T(\zeta-\mu)}{\partial\zeta}=\delta(\zeta-\mu)$.  The longitudinal resistivity $\rho_{xx}$ and Hall resistivity $\rho_{xy}$ are obtained through tensor inversions, 
\begin{align}
\rho_{xx}=\frac{\sigma_{xx}}{\sigma_{xx}^2+\sigma_{xy}^2}, \quad
\rho_{xy}=\frac{\sigma_{xy}}{\sigma_{xx}^2+\sigma_{xy}^2}. 
\end{align}

\begin{figure*}
	\includegraphics[width=16cm]{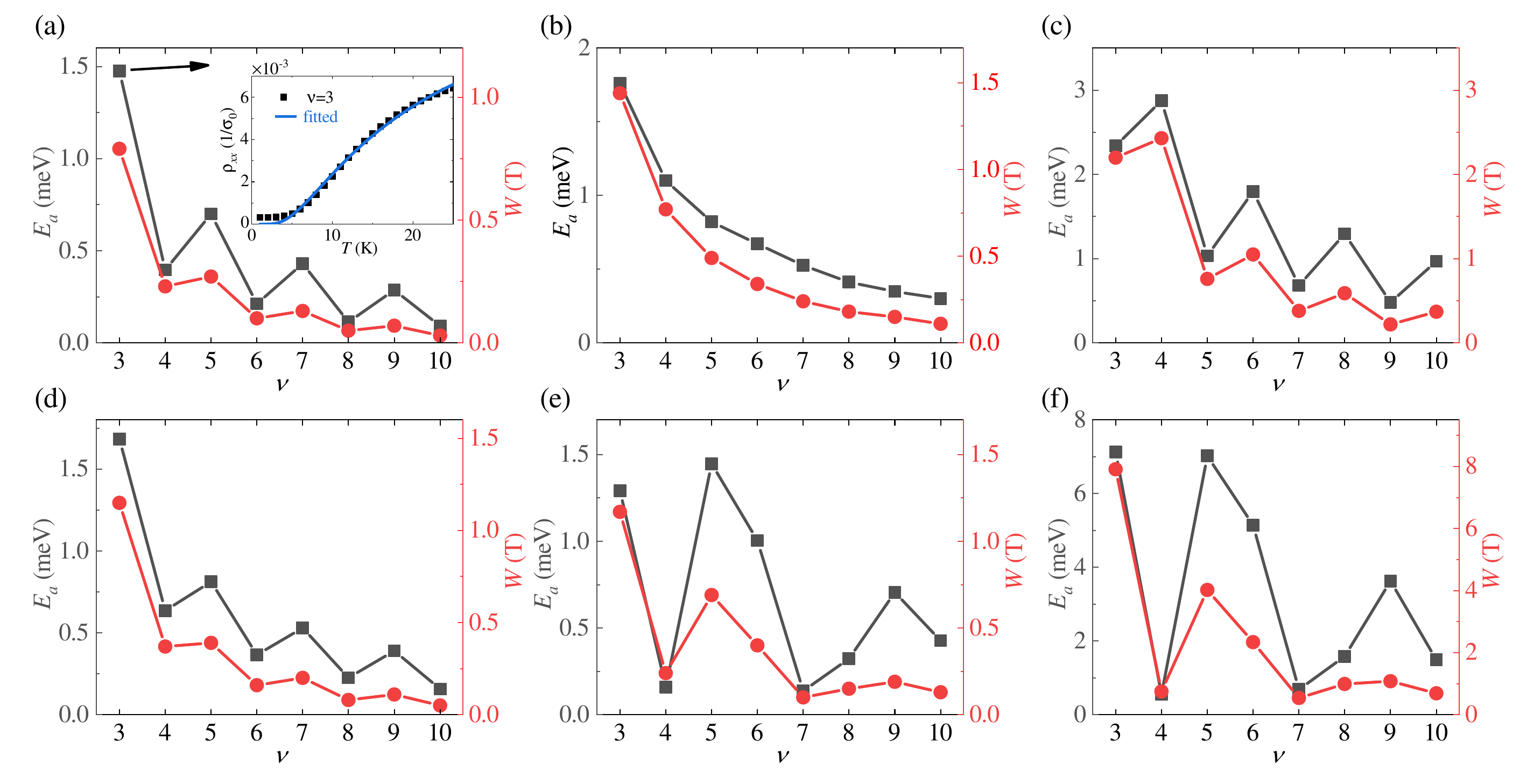}
	\caption{(Color online) The activation energy $E_a$ and plateau width $W$ (in unit of Tesla) versus the Hall plateau $\nu$.
	We choose $v_F=5\times10^5$ m/s, $\xi=600$ meV nm$^2$, and the other parameters as: 
	(a) $V_0=0$, $\Delta=20$ meV, $\mu=30$ meV; 
	(b) $V_0=0$, $\Delta=20$ meV, $\mu=45$ meV; 
	(c) $V_0=0$, $\Delta=20$ meV, $\mu=70$ meV;
	(d) $V_0=0$, $\Delta=-20$ meV, $\mu=40$ meV; 
	(e) $V_0=10$ meV, $\Delta=20$ meV, $\mu=42$ meV; 
	and (f) $V_0=35$ meV, $\Delta=20$ meV, $\mu=113$ meV. 
	The legends are the same in all figures.  
	The inset in (a) shows a fitting example at $\nu=3$, where the Arrhenius-type fitting formula is given as $\rho_{xx}=0.01312 e^{-\frac{1.4756}{k_BT}}$. }
	\label{Fig4}
\end{figure*}

\section{features of 2D quantum Hall effect} 

In Fig.~\ref{Fig3}, the magnetoconductivities and magnetoresisitivities of Cd$_3$As$_2$ thin films are plotted as functions of the magnetic field $B$ (or inverse magnetic field $B^{-1}$) for different temperatures.  We choose the same model parameters as Fig.~\ref{Fig1}(a), and set the chemical potential $\mu=30$ meV and the linewidth $\eta=0.1$ meV.  

At zero temperature $T=0$, in the magnetoconductivities, the features of 2D QHE can be captured: (i) $\sigma_{xx}$ shows Shubnikov-de Hass (SdH) oscillations with $B^{-1}$ and $\sigma_{xy}$ exhibits quantized plateaus; (ii) the peaks in $\sigma_{xx}$ correspond to the plateau transitions in $\sigma_{xy}$, which is caused by the magnetic field-driven LL crossing over the chemical potential, and the dips (minima) in $\sigma_{xx}$ correspond to the plateau centers in $\sigma_{xy}$. 
As $\sigma_{xx}\ll\sigma_{xy}$, we have the magnetoresistivities $\rho_{xx}=\sigma_{xx}\sigma_{xy}^{-2}$ and $\rho_{xy}=\sigma_{xy}^{-1}$, thus the features of 2D QHE can also be found in the magnetoresistivities. 

Moreover, the oscillation period in $\sigma_{xx}$ (or $\rho_{xx}$) is extracted as $\Delta_{1/B}=0.376$ T$^{-1}$.  This is understood from the Onsager's relation that the oscillation period $\Delta_{1/B}^O$ is inversely proportional to the Fermi surface area $S_F$~\cite{D.Shoeberg, L.Onsager}, 
\begin{align}
\Delta_{1/B}^O=\frac{2\pi e}{S_F}. 
\end{align}
Under no magnetic field, the Fermi surface has a circular shape and its area is calculated to be $S_F=0.02565$ nm$^{-2}$.  Then we obtain $\Delta_{1/B}^O=0.376$ T$^{-1}$, agreeing well with the extracted $\Delta_{1/B}$. 

Next we study the effect of finite temperatures in the magnetotransport properties.  As shown in Fig.~\ref{Fig3}, with increasing $T$, the thermal fluctuations will smoothen the SdH oscillations in $\sigma_{xx}$ and $\rho_{xx}$, and break the plateaus in $\sigma_{xy}$ and $\rho_{xy}$ as well. 
Explicitly, the quantum Hall insulating states are broken successively from the plateau edges to the centers, suggesting that the electronic states at the plateau edges are more fragile to the thermal fluctuations than those at the centers.  This observation is crucial in understanding the activation energy behavior discussed in next section. 

At a high temperature $T=50$ K, the magnetoconductivities and magnetoresistivities all exhibit smooth variations with $B$,
indicating that the quantum Hall states are broken completely by such a large thermal fluctuation.  
Correspondingly, the system is driven into the diffusive metal phase.  In Fig.~\ref{Fig3}(b) inset, we see that $\sigma_{xy}$ exhibits an almost linear dependence on the inverse magnetic field $B^{-1}$, which can be fitted by $\sigma_{xy}=5.63 B^{-0.98}$.  Note that in the 2D surface states, such a linear dependence of $\sigma_{xy}$ is caused by the thermal fluctuations and is different from the classical linear relationship $\sigma_{xy}\propto B^{-1}$ in a 3D Dirac semimetal at zero temperature~\cite{Abrikosov, Z.Cai, Y.X.Wang2023}.

\section{Activation energy} 

To further explore the effect of temperature in the magnetotransport, we calculate the activation energy $E_a$ that is associated with the dips of the longitudinal resistivity $\rho_{xx}$.  At each dip, $E_a$ is determined from the dependence of $\rho_{xx}$ on temperature.  Numerically, $E_a$ can be extracted from the Arrhenius-type fitting formula, 
\begin{align}
\rho_{xx}(T)=\rho_\infty e^{-\frac{E_a}{k_B T}}. 
\end{align}
In the calculations, the temperature is chosen from $T=1$ K up to that $\rho_{xx}$ reaches its maximum, with the interval $\Delta T=1$ K.  
Experimentally, the activation energy can be used to assess the energy scale of quantum states, \textit{e.g.}, the pressure-induced breakdown of 3D Dirac semimetal state in Cd$_3$As$_2$~\cite{S.Zhang} and the correlation-induced quantum anomalous Hall state in twisted bilayer graphene~\cite{M.Serlin}. 
In Fig.~\ref{Fig4}(a) inset, we present an example of calculating $E_a$ at $\nu=3$: when $B=1.87$ T, the zero-temperature $\rho_{xx}$ takes its minimum and the finite-temperature $\rho_{xx}(T)$ is fitted as $\rho_{xx}(T)=0.01312 e^{-\frac{1.4756}{k_BT}}$, from which the activation energy is extracted as $E_a=1.4756$ meV. 

In Fig.~\ref{Fig4}, the activation energy $E_a$ and plateau width $W$ are displayed as functions of the Hall plateau $\nu$ for different cases.  For convenience, we use Tesla as the unit of the plateau width.  
When the SIS breaking parameter $V_0=0$ and the model parameters are the same as Fig.~\ref{Fig1}(a), we see that (i) in Fig.~\ref{Fig4}(a) with a lower chemical potential $\mu=30$ meV, $E_a$ exhibits a sawtooth dependence on $\nu$, where $E_a$ at odd plateaus are larger than those at neighboring even plateaus;  
(ii) in Fig.~\ref{Fig4}(b) with an intermediate $\mu=45$ meV, $E_a$ does not show a sawtooth dependence, but decreases steadily with $\nu$, which is consistent with most 2D electron gas systems (e.g. see graphene in Appendix); 
(iii) in Fig.~\ref{Fig4}(c) with a higher $\mu=70$ meV, $E_a$ again exhibits a sawtooth dependence on $\nu$.  But now $E_a$ at odd plateaus are weaker than those at neighboring even plateaus, which is just opposite to case (i). 

\begin{figure} 
	\includegraphics[width=9.2cm]{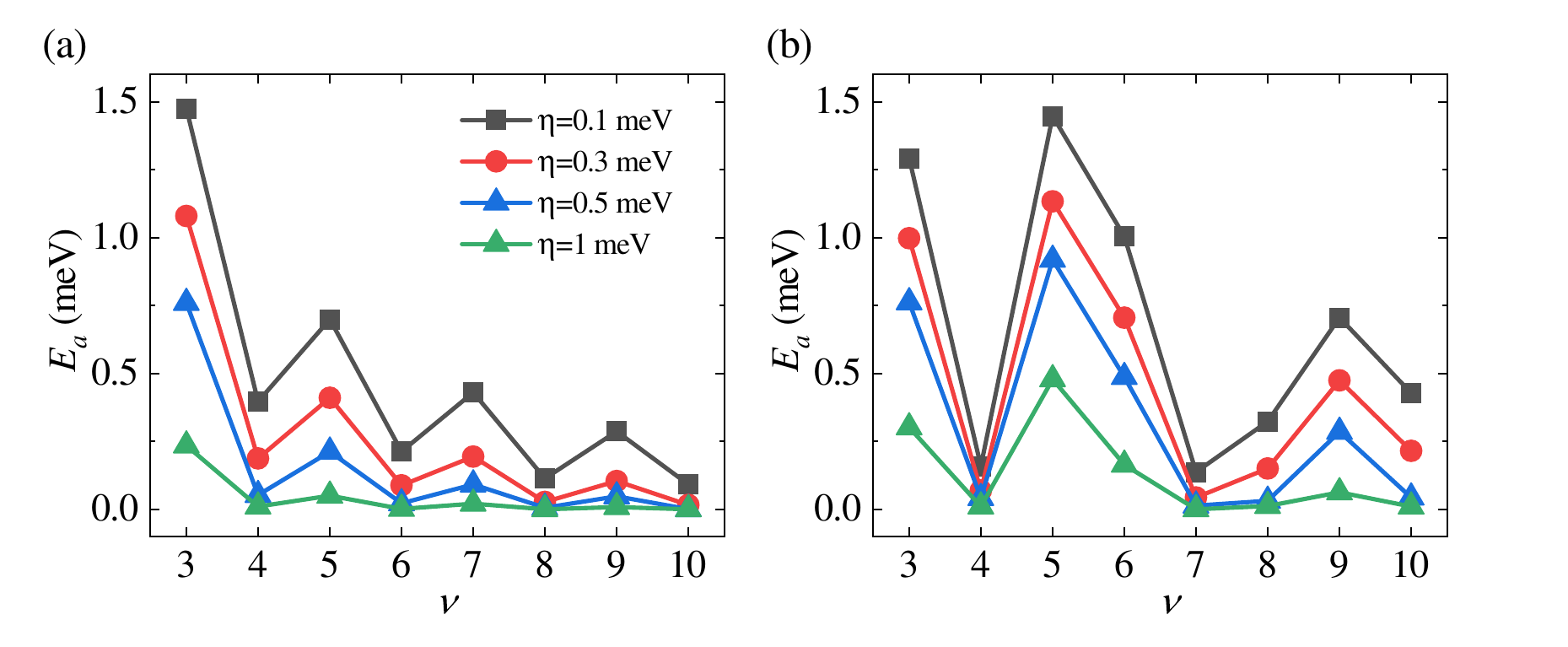}
	\caption{(Color online) The activation energy $E_a$ versus the Hall plateau $\nu$ for different linewidth $\eta$.  The parameters in (a) and (b) are the same as those in Figs.~\ref{Fig4}(a) and (e), respectively.  The legends are the same in both figures. } 
	\label{Fig5} 
\end{figure} 

We observe that the activation energy behavior is closely connected to the plateau width, since their variations are consistent with each other.  
As shown in the LL spectra of Fig.~\ref{Fig1}(a), the LLs originated from the top and bottom surface are independent and exhibit distinct evolutions with the magnetic field, resulting in the varying Hall plateau width with the chemical potential $\mu$.  
At a lower $\mu$, we see that $(n++)$ and $(n+-)$ LL branches are close and pair with each other, meaning that odd plateaus that lie between the LLs of the opposite parities are wider than the neighboring even plateaus between the LLs of the same parities. 
As a result, a higher temperature is needed to drive the dip insulating states of odd plateaus to the diffusive states than those of even plateaus, which  leads to a larger $E_a$ at odd plateaus.  This case is also seen in  Fig.~\ref{Fig4}(d), where $\mu=40$ meV and the model parameters are the same as Fig.~\ref{Fig1}(b).  
With increasing $\mu$, $(n+-)$ LL branch no longer pair with $(n++)$ LL branch, but will flow to $(n+1,++)$ LL branch.  
Consequently, at an intermediate $\mu$, the plateau width decreases steadily with $\nu$, leading to the monotonously decreasing of $E_a$.  
At a higher $\mu$, odd plateaus will be narrower than even plateaus, thus $E_a$ at odd plateaus are weaker than those at even plateaus.  

When the SIS breaking parameter $V_0$ is finite, the induced LL crossings will change the Hall plateau width and $E_a$ exhibits a nonmonotonic variation with $\nu$.  But in some specific cases, the behavior that $E_a$ at odd plateaus are larger than those at even plateaus can still be found. 
For example, this is seen in Fig.~\ref{Fig4}(e), where $\mu=42$ meV and the model parameters are the same as Fig.~\ref{Fig1}(c), and in Fig.~\ref{Fig4}(f), where $\mu=113$ meV and the model parameters are the same as Fig.~\ref{Fig1}(d). 
Moreover, in Fig.~\ref{Fig4}(e), we observe that $E_a$ at $\nu=7$ is close to zero, and in Fig.~\ref{Fig4}(f), $E_a$ at $\nu=4$ and $\nu=7$ are also close to zero.  Correspondingly, the nearly vanishing plateaus can be found.  According to these results, we point out that the activation energy is determined by the Hall plateau width, no matter whether the SIS breaking is present or not.  

We also study the effect of the impurity scatterings on the activation energy.  In the calculations, the impurity scatterings are represented by the linewidth $\eta$.  
In Figs.~\ref{Fig5}(a) and~\ref{Fig5}(b), we see that with increasing $\eta$, $E_a$ will get reduced and reach several hundreds of $\mu$eV, which can be compared with the experimental results in magnitude~\cite{D.A.Kealhofer}.  
At $\eta=1$ meV, $E_a$ at some plateaus is strongly suppressed and even reduced to zero, indicating that the Hall plateaus is completely broken by the impurity scatterings.  Therefore, we suggest that the SIS breaking can drive some Hall plateau become much narrower and the impurity scatterings further suppress the Hall plateau; their combined effects would lead to the Hall plateau absence that was observed in the experiment~\cite{D.A.Kealhofer}.

\section{Discussions and Conclusions}

To summarize, in this paper, we investigate the magnetotransport properties of the surface states in Cd$_3$As$_2$ thin films, focusing on the features of 2D QHE and activation energy.  We reveal that the activation energy is closely related to the Hall plateau width and the observed difference between odd and even plateaus in the experiment~\cite{D.A.Kealhofer} is due to the interplay of such essential factors: the chemical potential, the SIS breaking, and the impurity scatterings. 
Instead of the intersurface Coulomb interactions, which we admit must have certain influences on the activation energy, making it enhance or weaken, here we demonstrate that the noninteracting single-particle picture can also be used to understand the activation energy behavior, just as it has been widely used to explain many experimental results of topological materials~\cite{Y.Jiang, Y.F.Zhao, M.Mogi}.  

Since manipulating the surface states of the topological semimetals has become feasible, more states that cannot be obtained from the bulk materials are expected in the thin film configurations.  
Therefore, they will provide new opportunities to explore the exotic topological phases and thus require more theoretical and experimental studies in the future.

\section{Acknowledgments} 

This work was supported by the Natural Science Foundation of China (Grant No. 11804122, No. 11905054, No. 12275075), the National Key Research and Development Program of Ministry of Science and Technology (Grant No. 2021YFA1200700), the Fundamental Research Funds for the Central Universities from China, and the China Postdoctoral Science Foundation (Grant No. 2021M690970).

\section{Appendix: Activation energy in graphene} 

\begin{figure} 
	\includegraphics[width=9.2cm]{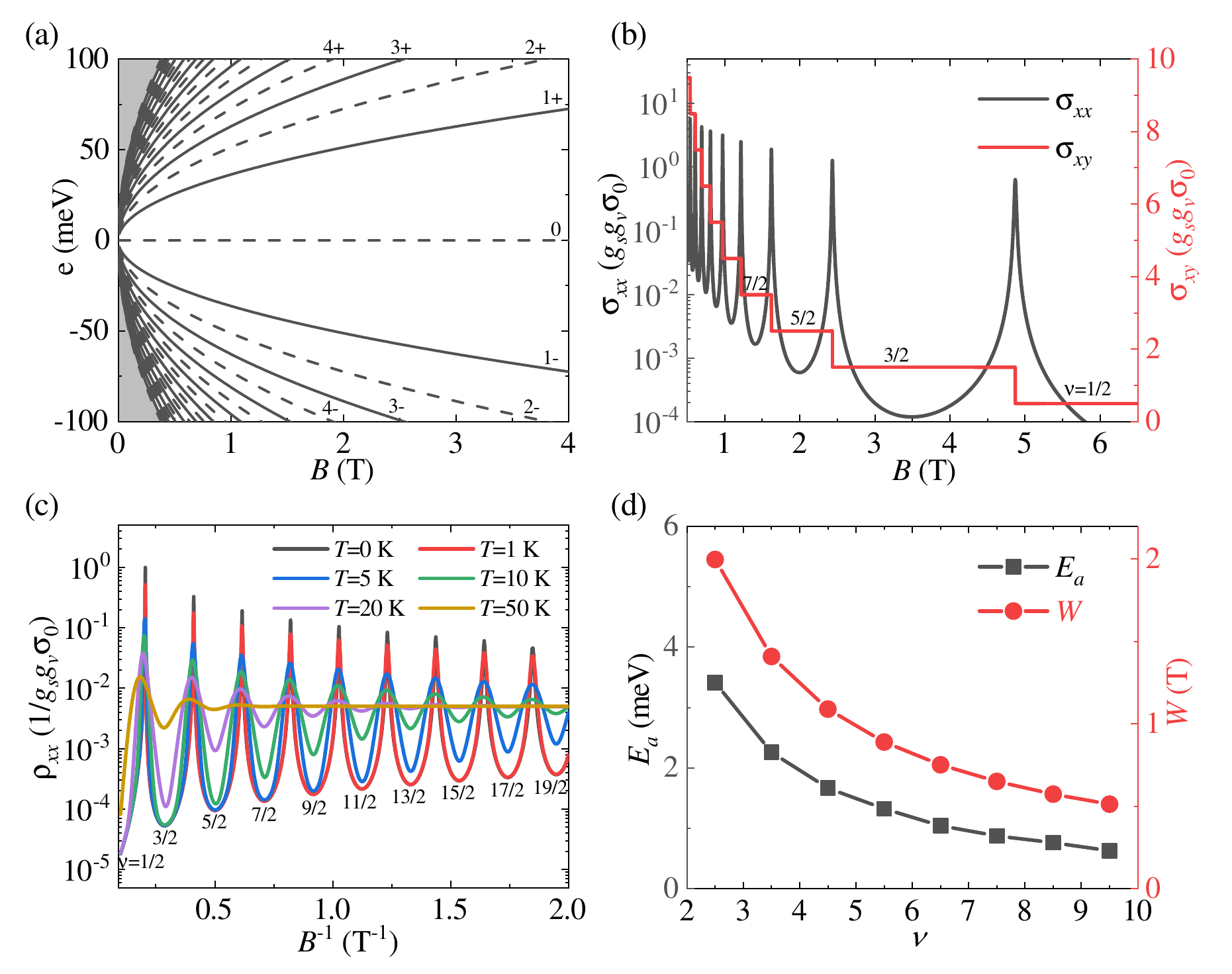} 
	\caption{(Color online) (a) The LL spectra in graphene, with the index labeled as  $(ns)$.  The even (odd) parity is represented by the solid (dashed) line.  (b) The zero-temperature longitudinal conductivity $\sigma_{xx}$ and Hall conductivity $\sigma_{xy}$ versus the magnetic field $B$.  (c) The longitudinal resistivity $\rho_{xx}$ versus the inverse magnetic field $B^{-1}$ for different temperatures $T$.  (d) The activation energy $E_a$ and the plateau width $W$ versus the Hall plateau $\nu$.  The half-integer Hall plateaus $\nu$ are labeled in (b) and (c).  In (b)-(d), we choose the chemical potential $\mu=80$ meV and the linewidth $\eta=0.1$ meV.} 
	\label{FigA1}
\end{figure} 

Here to illustrate the activation energy behavior in conventional 2D electron gas systems, we take graphene as an example.  In graphene, the single-particle states around one of the two inequivalent valleys are described by the Hamiltonian~\cite{A.H.Castro, M.O.Goerbig} 
\begin{align}
\hat H_{\tau_z}(\boldsymbol k)=v (\tau_z k_x\sigma_x+k_y\sigma_y), 
\tag{A1}
\end{align}
where $v=10^6$ m/s is the Fermi velocity, $\sigma$ is the Pauli matrice acting on the two sublattices, and $\tau_z=\pm1$ denotes the valley $\boldsymbol K/\boldsymbol K'$.  

Under a uniform magnetic field $\boldsymbol B=B\boldsymbol e_z$, the energies for the $n=0$ and $n\geq1$ LLs are solved as 
\begin{align*}
\varepsilon_0=0, \quad
\varepsilon_{ns}=s\sqrt{n}\omega, 
\tag{A2}
\end{align*}
respectively, where $s=\pm1$ denotes the conduction/valence band and $\omega=\frac{\sqrt2 v}{l_B}$ is the characteristic frequency.  Each LL owns fourfold degeneracy, including twofold spin degeneracy $g_s=2$ and twofold valley degeneracy $g_v=2$.  
We see that $\varepsilon_{ns}$ is proportional to $\sqrt B$, and the LL spacings decrease continuously with the index $n$, as shown in Fig.~\ref{FigA1}(a) of the LL spectra.  
Since the parity symmetry with the operator $\hat{\cal P}=(-1)^{\hat a^\dagger \hat a}\sigma_z$ is present in graphene, each LL carries a definite even or odd parity, as represented by the solid or dashed lines.  Note that the parity property of the LLs in valley $\boldsymbol K'$ is just opposite to that in valley $\boldsymbol K$. 

With the help of the Kubo-Bastin formula, at zero temperature, the longitudinal conductivity $\sigma_{xx}$ and Hall conductivity $\sigma_{xy}$ are derived as
\begin{align}
&\sigma_{xx}(T=0)=g_sg_v\sigma_0\frac{v^2\eta^2}{2\pi l_B^2} \sum_{n\geq 0}\sum_{s,s'}  
\frac{1}{(\mu-\varepsilon_{ns})^2+\eta^2}
\nonumber\\
&\quad\times\frac{1}{(\mu-\varepsilon_{n+1,s'})^2+\eta^2}, 
\tag{A3} 
\end{align}
and
\begin{align}
&\sigma_{xy}(T=0)=g_sg_v\sigma_0\frac{v^2}{2l_B^2}\sum_{n\geq 0}\sum_{s,s'}
\frac{(\varepsilon_{n+1,s'}-\varepsilon_{ns})^2-\eta^2}
{[(\varepsilon_{n+1,s'}-\varepsilon_{ns})^2+\eta^2]^2}
\nonumber\\
&\times\big[\theta(\mu-\varepsilon_{ns})\theta(\varepsilon_{n+1,s'}-\mu)
-\theta(\varepsilon_{ns}-\mu)\theta(\mu-\varepsilon_{n+1,s'})\big], 
\tag{A4}
\label{A4}
\end{align}
respectively.  In Eq.~(\ref{A4}), when the chemical potential is positive $\mu>0$ and there are no impurity scatterings $\eta=0$, we have 
\begin{align}
&\sigma_{xy}(T=0)=g_sg_v\sigma_0\frac{v^2}{2l_B^2}
\nonumber\\
&\times\sum_{n\geq 0}\Big[\frac{1}{(\varepsilon_{n+1,+}-\varepsilon_{n+})^2}
+\frac{1}{(\varepsilon_{n+1,+}-\varepsilon_{n-})^2}\Big]\theta(\mu-\varepsilon_{n+})
\nonumber\\
&=g_sg_v\sigma_0\sum_{n\geq 0}\big(n+\frac{1}{2}\big)\theta(\mu-\varepsilon_{n+}), 
\tag{A5} 
\label{A5} 
\end{align}
which retrieves the famous half-integer quantum Hall effect of massless Dirac fermions in graphene~\cite{K.S.Novoselov, Yuanbo.Zhang, V.P.Gusynin}.  
The finite-temperature conductivities are calculated by using Eq.~(\ref{finiteT}) in the main text. 

In the calculations, we choose the chemical potential $\mu=80$ meV and the linewidth $\eta=0.1$ meV.  In Fig.~\ref{FigA1}(b), the zero-temperature conductivities $\sigma_{xx}$ and $\sigma_{xy}$ are plotted, where the features of 2D QHE can be observed, including the SdH oscillations in $\sigma_{xx}$ and the quantized plateaus in $\sigma_{xy}$.  In Fig.~\ref{FigA1}(c), the longitudinal resistivity $\rho_{xx}$ is plotted for different temperatures $T$.  With increasing $T$, the SdH oscillations are smeared out and $\rho_{xx}$ becomes a smooth curve.   

Finally, the results of the activation energy $E_a$ and plateau width $W$ are plotted in Fig.~\ref{FigA1}(d).  We see that $E_a$ decreases steadily with $\nu$, which is consistent with the variation of $W$, supporting the results of Fig.~\ref{Fig4}(b) in the main text.  As the qualitative relations of the Hall plateau width remain unchanged to the varying  chemical potential, such an activation energy behavior is believed to always hold in graphene.


\begin{thebibliography}{100}

\bibitem{M.Z.Hasan} 
M. Z. Hasan and C. L. Kane, 
Colloquium: Topological insulators, 
Rev. Mod. Phys. 82 (2010) 3045. 

\bibitem{X.L.Qi}
X. L. Qi and S. C. Zhang, 
Topological insulators and superconductors, 
Rev. Mod. Phys. 83 (2011) 1057. 

\bibitem{N.P.Armitage}
N. P. Armitage, E. J. Mele, and A. Vishwanath, 
Weyl and Dirac semimetals in three-Dimensional solids, 
Rev. Mod. Phys. 90 (2018) 015001.

\bibitem{B.Q.Lv}
B. Q. Lv, T. Qian, and H. Ding, 
Experimental perspective on three-dimensional topological semimetals, 
Rev. Mod. Phys. 93 (2021) 025002. 

\bibitem{X.G.Wan} 
X. G. Wan, A. M. Turner, A. Vishwanath, and S. Y. Savrasov,
Topological semimetal and Fermi-arc surface states in the electronic structure of pyrochlore iridates, Phys. Rev. B 83 (2011) 205101.

\bibitem{A.C.Potter}
A. C. Potter, I. Kimchi, and A. Vishwanath, 
Quantum oscillations from surface Fermi arcs in Weyl and Dirac semimetals, 
Nat. Commun. 5 (2014) 5161.

\bibitem{P.J.W.Moll}
P. J. W. Moll, N. L. Nair, T. Helm, A. C. Potter, I. Kimchi, A. Vishwanath, and J. G. Analytis, 
Transport evidence for Fermi-arc-mediated chirality transfer in the Dirac semimetal
Cd$_3$As$_2$, 
Nature (London) 535 (2016) 266.

\bibitem{M.Kargarian}
M. Kargarian, Y. M. Lu, and M. Randeria, 
Deformation and stability of surface states in Dirac semimetals, 
Phys. Rev. B 97, (2018) 165129.  

\bibitem{B.J.Yang}  
B. J. Yang and N. Nagaosa, 
Classification of stable three-dimensional Dirac semimetals with nontrivial topology
Nat.Commun. 5 (2014) 4898. 

\bibitem{Z.Wang} 
Z. Wang, H. Weng, Q. Wu, X. Dai, and Z. Fang, 
Three-dimensional Dirac semimetal and quantum transport in Cd$_3$As$_2$, 
Phys. Rev. B 88 (2013) 125427. 

\bibitem{S.Borisenko}
S. Borisenko, Q. Gibson, D. Evtushinsky, V. Zabolotnyy, B. B\"uchner, and R. J. Cava, 
Experimental realization of a three-dimensional Dirac semimetal, 
Phys. Rev. Lett 113 (2014) 027603. 

\bibitem{Z.K.Liu}
Z. K. Liu, J. Jiang, B. Zhou, Z. J.Wang, Y. Zhang, H. M. Weng, D. Prabhakaran, S. K. Mo,
H. Peng, P. Dudin, T. Kim, M. Hoesch, Z. Fang, X. Dai, Z. X. Shen, D. L. Feng, Z. Hussain, and Y. L. Chen, 
A  stable three-dimensional topological Dirac semimetal Cd$_3$As$_2$, 
Nat. Mat. 13 (2014) 677. 

\bibitem{S.Jeon}
S. Jeon, B. B. Zhou, A. Gyenis, B. E. Feldman, I. Kimchi, A. C. Potter, Q. D. Gibson, R. J. Cava, A. Vishwanath and A. Yazdani, 
Landau quantization and quasiparticle interference in the three-dimensional Dirac semimetal Cd$_3$As$_2$,
Nat. Mat. 13 (2014) 851. 

\bibitem{T.Liang}
T. Liang, Q. Gibson, M. N. Ali, M. Liu, R. J. Cava and N. P. Ong, 
Ultrahigh mobility and giant magnetoresistance in the Dirac semimetal Cd$_3$As$_2$, 
Nat. Mat. 14 (2015) 280. 

\bibitem{D.A.Kealhofer}
D. A. Kealhofer, L. Galletti, T. Schumann, A. Suslov, and S. Stemmer, 
Topological insulator state and collapse of the quantum Hall effect in a three-dimensional Dirac semimetal heterojunction, 
Phys. Rev. X 10 (2020) 011050. 

\bibitem{B.Guo}
B. Guo, A. C. Lygo, X. Dai, and S. Stemmer, 
$\nu=0$ quantum Hall state in a cadmium arsenide thin film, 
APL Mater. 10 (2022) 091116.

\bibitem{A.C.Lygo}
A. C. Lygo, B. Guo, A. Rashidi, V. Huang, P. Cuadros-Romero, and S. Stemmer, 
Two-dimensional topological insulator state in Cadmium Arsenide thin films, 
Phys. Rev. Lett. 130 (2023) 046201.  

\bibitem{M.Uchida}
M. Uchida, Y. Nakazawa, S. Nishihaya, K. Akiba, M. Kriener, Y. Kozuka, A. Miyake, Y. Taguchi, M. Tokunaga, N. Nagaosa, Y. Tokura, and M. Kawasaki, 
Quantum Hall states observed in thin films of Dirac semimetal Cd$_3$As$_2$, 
Nat. Commun. 8 (2017) 2274. 

\bibitem{T.Schumann}
T. Schumann, L. Galletti, D. A. Kealhofer, H. Kim, M. Goyal, and S. Stemmer, 
Observation of the quantum Hall effect in confined films of the three-dimensional Dirac semimetal Cd$_3$As$_2$, 
Phys. Rev. Lett. 120 (2018) 016801.

\bibitem{C.Zhang}
C. Zhang, Y. Zhang, X. Yuan, S. Lu, J. Zhang, A. Narayan, Y. Liu, H. Zhang, Z. Ni, R. Liu, E. S. Choi, A. Suslov, S. Sanvito, L. Pi, H. Z. Lu, A. C. Potter, and F. Xiu,	
Quantum Hall effect based on Weyl orbits in Cd$_3$As$_2$,  
Nature (London) 565 (2019) 331.

\bibitem{J.P.Eisenstein}
J. P. Eisenstein and A. H. MacDonald, 
Bose-Einstein condensation of excitons in bilayer electron systems, 
Nature (London) 432 (2004) 691.

\bibitem{D.Tilahun}
D. Tilahun, B. Lee, E. M. Hankiewicz, and A. H. MacDonald, 
Quantum Hall superfluids in topological insulator thin films, 
Phys. Rev. Lett. 107 (2011) 246401.

\bibitem{Y.Zhang} 
Y. Zhang, K. He, C. Z. Chang, C. L. Song, L. L. Wang, X. Chen, J. F. Jia, Z. Fang, X. Dai, W. Y. Shan, S. Q. Shen, Q. Niu, X. L. Qi, S. C. Zhang, X. C. Ma, and Q. K. Xue, 
Crossover of the three-dimensional topological insulator Bi$_2$Se$_3$ to the two-dimensional limit, 
Nat. Phys. 6 (2010) 584.

\bibitem{W.Y.Shan}
W. Y. Shan, H. Z. Lu, and S. Q. Shen, 
Effective continuous model for surface states and thin films of three-dimensional topological insulators,  
New J. Phys. 12 (2010) 043048.  

\bibitem{Y.X.Wang2023}
Y. X. Wang and Z. Cai, 
Quantum oscillations and three-dimensional quantum Hall effect in ZrTe$_5$, 
Phys. Rev. B 107 (2023) 125203.  

\bibitem{H.W.Wang}
H. W. Wang, B. Fu, and S. Q. Shen, 
Theory of the anomalous Hall effect in the transition metal pentatellurides ZrTe$_5$ and HfTe$_5$,
Phys. Rev. B 108 (2023) 045141.   

\bibitem{X.Xiao} 
X. Xiao, S. A. Yang, Z. Liu, H. Li, and G. Zhou,
Anisotropic quantum confinement effect and electric control of surface states in Dirac semimetal nanostructures, 
Sci. Rep. 5 (2015) 7898. 

\bibitem{H.Z.Lu}
H. Z. Lu, W. Y Shan, W. Yao, Q. Niu, and S. Q. Shen, 
Massive Dirac fermions and spin physics in an ultrathin film of topological insulator, 
Phys. Rev. B 81 (2010) 115407.  

\bibitem{T.Devakul}
T. Devakul, Y. H. Kwan, S. L. Sondhi, and S. A. Parameswaran, 
Quantum oscillations in the zeroth Landau level: serpentine Landau fan and the chiral anomaly, 
Phys. Rev. Lett. 127 (2021) 116602. 

\bibitem{Y.X.Wang2022}  
Y. X. Wang and F. Li, 
Unconventional optical selection rules in ZrTe$_5$ under an in-plane magnetic field, 
Phys. Rev. B 106 (2022) 205102. 

\bibitem{B.A.Bernevig2006b}
B. A. Bernevig, T. L. Hughes, and S. C. Zhang, 
Quantum spin Hall effect and topological phase transition in HgTe Quantum Wells,
Science 314 (2006) 1757. 

\bibitem{M.Konig} 
M. K\"onig, S. Wiedmann, C. Br\"une, A. Roth, H. Buhmann, L. W. Molenkamp, X. L. Qi, and S. C. Zhang, 
Quantum spin Hall insulator state in HgTe quantum wells, 
Science 318 (2007) 766.  

\bibitem{S.Wu}
S. Wu, V. Fatemi, Q. D. Gibson, K. Watanabe, T. Taniguchi, R. J. Cava, P. Jarillo-Herrero, 
Observation of the quantum spin Hall effect up to 100 kelvin in a monolayer crystal, 
Science 359 (2018) 76.  

\bibitem{S.Tang}
S. Tang, C. Zhang, D. Wong, Z. Pedramrazi, H. Z. Tsai, C. Jia, B. Moritz, M. Claassen, H. Ryu, S. Kahn, J. Jiang, H. Yan, M. Hashimoto, D. Lu, R. G. Moore, C. C. Hwang, C. Hwang, Z. Hussain, Y. Chen, M. M. Ugeda, Z. Liu, X. Xie, T. P. Devereaux, M. F. Crommie, S. K. Mo and Z. X. Shen,
Quantum spin Hall state in monolayer 1T’-WTe$_2$, 
Nat. Phys. 13 (2017) 683.  

\bibitem{C.L.Kane} 
C. L. Kane and E. J. Mele, 
Quantum spin Hall effect in graphene, 
Phys. Rev. Lett. 95 (2005) 226801.  

\bibitem{C.C.Liu}
C. C. Liu, W. Fen, and Y. Yao, 
Quantum spin Hall effect in silicene and two-dimensional germanium, 
Phys. Rev. Lett. 107 (2011) 076802.  

\bibitem{M.Ezawa}
M. Ezawa, 
Valley-polarized metals and quantum anomalous Hall effect in silicene, 
Phys. Rev. Lett. 109 (2012) 055502.  

\bibitem{X.Qian}
X. Qian, J. Liu, L. Fu, and J. Li, 
Quantum spin Hall effect in two-dimensional transition metal dichalcogenides, 
Science, 346 (2014) 1344.  

\bibitem{M.A.Cazalilla}
M. A. Cazalilla, H. Ochoa, and F. Guinea, 
Quantum spin Hall effect in two-dimensional crystals of transition-metal dichalcogenides
Phys. Rev. Lett. 113 (2014) 077201.

\bibitem{A.Bastin}
A. Bastin, C. Lewiner, O.Betbeder-Matibet, and P. Nozieres, 
Quantum oscillations of the Hall effect of a fermion gas with random impurity scattering, 
J. Phys. Chem. Solids. 32 (1971) 1811. 

\bibitem{L.Smrcka}
L. Smrcka and P. Streda, 
Transport coefficients in strong magnetic fields, 
J. Phys. C: Solid State Phys. 10 (1977) 2153. 

\bibitem{G.D.Mahan}
G. D. Mahan, 
\textit{Many-Particle Physics}, 3rd ed. (Plenum, New York, 2000). 

\bibitem{H.Watanabe2020}
H. Watanabe and Y. Yanase, 
Nonlinear electric transport in odd-parity magnetic multipole systems: Application to Mn-based compounds, 
Phys. Rev. Res. 2 (2020) 043081. 

\bibitem{H.Watanabe2021}
H. Watanabe and Y. Yanase, 
Chiral photocurrent in parity-violating magnet and enhanced responses in topological antiferromagnet, 
Phys. Rev. X 11 (2021) 011001. 

\bibitem{Z.Cai}
Z. Cai and Y. X. Wang, 
Magnetic field driven Lifshitz transition and one-dimensional Weyl nodes in three-dimensional pentatellurides, 
Phys. Rev. B 108 (2023) 155202.   

\bibitem{D.Shoeberg} 
D. Shoenberg, 
\textit{Magnetic Oscillations in Metals}, 
(Cambridge University Press, Cambridge, 1984). 

\bibitem{L.Onsager}
L. Onsager,
Interpretation of the de Haas-van Alphen effect, 
Philos. Mag. 43, (1952) 1006. 

\bibitem{Abrikosov}
A. A. Abrikosov, 
Quantum magnetoresistance, 
Phys. Rev. B 58 (1998) 2788.  

\bibitem{S.Zhang}
S. Zhang, Q. Wu, L. Schoop, M. N. Ali, Y. Shi, N. Ni, Q. Gibson, S. Jiang, V. Sidorov, W. Yi, J. Guo, Y. Zhou, D. Wu, P. Gao, D. Gu, C. Zhang, S. Jiang, K. Yang, A. Li, Y. Li, X. Li, J. Liu, X. Dai, Z. Fang, R. J. Cava, L. Sun, and Z. Zhao, 
Breakdown of three-dimensional Dirac semimetal state in pressurized Cd$_3$As$_2$,
Phys. Rev. B 91 (2015) 165133.  

\bibitem{M.Serlin}
M. Serlin, C. L. Tschirhart, H. Polshyn, Y. Zhang, J. Zhu, K. Watanabe, T. Taniguchi, L. Balents, and A. F. Young, 
Intrinsic quantized anomalous Hall effect in a moir\'e heterostructure, 
Science 367 (2020) 900.  

\bibitem{Y.Jiang}  
Y. Jiang, J. Wang, T. Zhao, Z. L. Dun, Q. Huang, X. S. Wu, M. Mourigal, H. D. Zhou, W. Pan, M. Ozerov, D. Smirnov, and Z. Jiang, 
Unraveling the topological phase of ZrTe$_5$ via magnetoinfrared spectroscopy, 
Phys. Rev. Lett. 125, (2020) 046403. 

\bibitem{Y.F.Zhao}
Y. F. Zhao, R. Zhang, R. Mei, L. Zhou, H. Yi, Y. Q. Zhang, J. Yu, R. Xiao, K. Wang, N. Samarth, M. H. W. Chan, C. X. Liu, and C. Z. Chang,
Tuning the Chern number in quantum anomalous Hall insulators, 
Nature (London) 588 (2020) 419. 

\bibitem{M.Mogi}
M. Mogi, Y. Okamura, M. Kawamura, R. Yoshimi, K. Yasuda, A. Tsukazaki, K. S. Takahashi, T. Morimoto, N. Nagaosa, M. Kawasaki, Y. Takahashi and Y. Tokura,
Experimental signature of the parity anomaly in a semi-magnetic topological insulator, 
Nat. Phys. 18 (2022) 390.  

\bibitem{A.H.Castro}
A. H. Castro Neto, F. Guinea, N. M. R. Peres, K. S. Novoselov, and A. K. Geim, 
The electronic properties of graphene,  
Rev. Mod. Phys. 81 (2009) 109.   	

\bibitem{M.O.Goerbig}
M. O. Goerbig, 
Electronic properties of graphene in a strong magnetic field,  
Rev. Mod. Phys. 83 (2011) 1193.  

\bibitem{K.S.Novoselov}
K. S. Novoselov, A. K. Geim, S. V. Morozov, D. Jiang, M. I. Katsnelson, I. V. Grigorieva, S. V. Dubonos, and A. A. Firsov, 
Two-dimensional gas of massless Dirac fermions in graphene, 
Nature (London) 438 (2005) 197.  

\bibitem{Yuanbo.Zhang} 
Y. Zhang, Y. W. Tan, H. L. Stormer, and P. Kim, 
Experimental observation of the quantum Hall effect and Berry’s phase in graphene, 
Nature (London) 438 (2005) 201.  

\bibitem{V.P.Gusynin}
V. P. Gusynin and S. G. Sharapov, 
Transport of Dirac quasiparticles in graphene: Hall and optical conductivities, 
Phys. Rev. B 73 (2006) 245411. 

\end{thebibliography}
\end{document}